\documentclass[preprint,aps,showpacs,floatfix]{revtex4}
\usepackage{amsmath,amssymb,graphicx,epsfig}

\newcommand{\ii}{{\bf {i}}}
\newcommand{\jj}{{\bf {j}}}
\newcommand{\vv}{{\bf {v}}}
\newcommand{\rr}{{\bf {r}}}
\newcommand{\pp}{{\bf {p}}}

\newcommand{\RR}{{\bf {R}}}
\newcommand{\VV}{{\bf {V}}}

\newcommand{\ep}{{\mbox{$\epsilon_p$}}}
\newcommand{\eF}{{\mbox{$\epsilon_F$}}}
\newcommand{\vF}{{\mbox{$v_F$}}}
\newcommand{\pF}{{\mbox{$p_F$}}}
\newcommand{\boldnabla}{{\mbox{\boldmath $\nabla$}}}

\begin{document}
%\draft
%\def\pd#1#2{\frac{\partial #1}{\partial #2}}

\title{\bf 
Thermal flux of
quasiparticles and the transition between two regimes of
turbulence in $^3$He-B}

\author{C. F. Barenghi${}^1$, Y. A. Sergeev${}^2$, 
N. Suramlishvili${}^{1,3}$, and P. J. van Dijk${}^2$}

\affiliation {
${}^1$School of Mathematics and Statistics,
Newcastle University, Newcastle upon Tyne, NE1 7RU,
${}^2$School of Mechanical and Systems Engineering,
Newcastle University, Newcastle upon Tyne, NE1 7RU,
${}^3$Andronikashvili Institute of Physics, Tbilisi,
0177, Georgia}
\date {\today}

\begin {abstract}

We compute the Andreev reflection coefficient of a flux of thermal
quasiparticles in $^3$He-B incident upon various two-dimensional
vortex configurations. We find that, for the same number of positive
and negative vortex
points, the reflection coefficient is much reduced if the points
are arranged in pairs (which corresponds to a gas of vortex rings
in three-dimensions) rather than random (which corresponds to
a vortex tangle in three-dimensions). This results is consistent
with measurements performed by Bradley {\it et al.} at the
University of Lancaster.

\end{abstract}

\maketitle

Experimental and theoretical studies
of turbulence in superfluid $^4$He and $^3$He 
have revealed features
which are similar to what happens in the 
classical turbulence but also
features which have no classical counterpart. 
The study of quantum turbulence at very low temperatures is particularly
interesting because, in the absence of the normal fluid, the turbulence
consists simply of a tangle of reconnecting, thin vortex filaments. 
Since all filaments have the same core radius and the same circulation,
the problem of turbulence is reduced to the geometry and the topology of the
tangle. Unfortunately, unlike what happens in the study of ordinary
turbulence, only few methods of visualization are possible in liquid
helium. In superfluid $^3$He-B, the Andreev scattering technique can be
used to detect vortex filaments. This technique, pioneered at Lancaster,
is based on the fact that the
energy dispersion curve, $E=E({\bf p})$ of quasiparticles of momentum
$\bf p$ is tied to the reference frame of the superfluid, so, 
in a superfluid moving with
velocity ${\bf v}_s$, the dispersion curve becomes $E({\bf p})+
{\bf p} \cdot {\bf v}_s$ (see the review article~\cite{Fisher2008}).
Thus a side of a vortex line presents a
potential barrier to oncoming quasiparticles, which can be
reflected back almost exactly becoming quasiholes; the other side
of the vortex lets the quasiparticles to go through. Quasiholes
are reflected or transmitted in the opposite way. The vortex thus
casts a symmetric shadow for quasiparticles at one side and
for quasiholes at the other side~\cite{Nugzar_1}, 
and, by measuring the flux of
excitations, one detects the presence of the vortex.

However the extrapolation from the scattering of quasiparticles
off one vortex to the scattering off many vortices may be nontrivial.
In our earlier work~\cite{Nugzar_2} we used a simplified
2-dimensional vortex point model, and found that the Andreev shadow
caused by simple configurations of several point vortices is not
necessarily equal to the sum of shadows of individual, isolated
vortices (in the cited work such a phenomenon was called a `partial
screening'). This result may have non-trivial implications for
the interpretation of Andreev reflection measurements of the vortex
line density in turbulent $^3$He-B.

The present work is motivated by the experimental observations of Bradley
{\it et al.}~\cite{BradleyPRL2005} of the transition from a
gas of vortex rings to a dense vortex tangle. Still using the simplified
2-dimensional model, the question which we ask is whether a gas of
vortex rings and a dense tangle have a different Andreev signature.

We start with describing the motion of vortices which, in the
considered two-dimensional approximation, become vortex points.
Such a
two-dimensional system of point vortices in the inviscid fluid 
is known as the Onsager's point vortex gas.
Each vortex point moves with the flow field generated by all other
vortices. In the $(x,\,y)$-plane, the $i^{\rm th}$ vortex point
located at time $t$ at the position
${\rr}_i(t)=x_i(t)\ii+y_i(t)\jj$, where $\ii$ and $\jj$ are,
respectively, the unit vectors in the $x$- and $y$-directions,
generates the fluid velocity field
\begin{equation}
{\vv}_i({\rr},\,t)=\frac{\kappa_i}{2\pi|{\rr}-{\rr}_i|^2}
[-(y-y_i)\ii+(x-x_i)\jj], \label{eq:vel_vortex}
\end{equation}
where $\kappa_i$ is the circulation generated by the $i^{\rm th}$
vortex. In the considered two-dimensional point vortex gas
modelling turbulent $^3$He-B, $\kappa_i=\pm \kappa$ with 
$\kappa=\pi\hbar/m
\approx0.662\times
10^{-3}~\textrm{cm}^2/\textrm{s}$,
where $\kappa$ is the quantum of circulation in $^3$He-B, $m$ is
the mass of $^3$He atom, and the signs plus and minus denote
respectively a vortex generating the anticlockwise rotation of the
fluid in the $(x,\,y)$-plane, and an antivortex generating the
clockwise rotation. At the point $\rr $, the velocity field
created by the system of $N$ vortices is given by the superposition of
the velocity fields generated by all vortices.

An isolated pair of a vortex and an antivortex can be considered as a
two-dimensional model of a three-dimensional vortex ring. Such a pair
moves through the fluid, in the direction orthogonal to the line
connecting the vortex and the antivortex, with the velocity
$V=\kappa/(2\pi d)$,
where $d$ is the distance between the vortex points in this pair.
Obviously, in the point vortex gas a vortex and an antivortex can be
considered a pair (a 2D vortex ring) if $d$ is much smaller than the
average distance between the center of this pair and the nearest
vortex which does not belong to this pair.

To model turbulence in $^3$He-B, we consider the following three
configurations of the vortex points:

\noindent$1^{\rm o}$. A spatially
random system of $N$ point vortices of the same
polarity; such a system has net circulation $\vert \kappa_t \vert=N\kappa$.
This configuration can be considered as the two-dimensional model of
a polarized vortex bundle~\cite{Alamri}.

\noindent$2^{\rm o}$. A random, statistically uniform system of $N/2$
vortices and $N/2$ antivortices in the case where the distance between
any two vortex points is not much smaller than $L/\sqrt{N}$, where $L$
is the size of the computational domain, so that vortices and
antivortices are not organized in pairs.
In such a system there is no net circulation, $\kappa_t=0$. This
configuration can be considered as a two-dimensional model of a
vortex tangle.

\noindent$3^{\rm o}$. A system of vortex-antivortex pairs
such that the distance $d$
between the vortex points in each pair is the same, but locations and
orientations of pairs are random. This configuration can be considered the
two-dimensional model of a gas of vortex rings.

The configurations $2^{\rm o}$ and $3^{\rm o}$ can be characterized by
two geometric quantities, $d$ and the average distance, say $a$
between the centers of two neighbouring vortex-antivortex pairs. The
relevant non-dimensional parameter characterizing the configuration
is $\zeta=d/a\leq1$.
Gas of small, compared to the mean intervortex distance, 
vortex-antivortex pairs (configuration $3^{\rm o}$) corresponds to 
$\zeta\ll1$.
Of particular interest is the transition
between configurations $3^{\rm o}$ and $2^{\rm o}$. Such a transition
corresponds, in a modelling sense, to the experimental observations of
Bradley {\it et al.}~\cite{BradleyPRL2005} of the transition from the
gas of vortex rings to the dense vortex tangle. In the considered
two-dimensional system of vortex points this transition can be modelled
by increasing the parameter $\zeta$ from small values corresponding to $d$
of the order of $100\xi_0$, where $\xi_0\approx0.85\times10^{-5}\,{\rm
cm}$ is the zero-temperature coherence length, to values of the
order of unity.

We consider the two-dimensional problem of ballistic
propagation of quasiparticle excitations in the flow field of the
point vortex gas and the gas of vortex-antivortex pairs described
by
configurations $1^{\rm o}$, $2^{\rm o}$, and $3^{\rm o}$.
Neglecting spatial variations of the order parameter, in the presence
of the flow field the energy
of thermal excitation can be written as
\begin{equation}
E=\sqrt{\ep^2+\Delta_0^2}+\pp \cdot \vv_s(\rr,\,t), \label{eq:E}
\end{equation}
where $\ep=p^2/(2m^*)-\eF$
is the kinetic energy of a thermal excitation of momentum $\pp$
relative to the Fermi
energy $\eF\approx 2.27 \times 10^{-16}~ \rm erg$
(here and below the numerical values are taken
at zero bar pressure~\cite{Wheatley}),
$m^* \approx 3.01 m =1.51 \times 10^{-23}~\rm g$,
with $m$ being the mass of the $^3$He atom, and
$\Delta_0=1.76k_B T_c \approx 2.43 \times
10^{-19}~\rm erg$, with $k_B$ being the Boltzmann's constant and $T_c$
the
critical temperature, is the superfluid energy gap. (It should be
noted that the superfluid energy gap
was assumed a constant value, $\Delta_0$.
because we are concerned with the behaviour of thermal
excitation at distances $r$ from the vortex core larger than the
zero-temperature coherence length $\xi_0$.)  Excitations with
$\ep>0$ are called quasiparticles, and excitations with
$\ep<0$ are called quasiholes.

Following the approach of Refs.~\cite{Nugzar_1,Nugzar_2} we assume
that the interaction term $\pp \cdot \vv_s$ varies on a
spatial scale which is larger than $\xi_0$ so that the excitation
can be regarded as a compact object of momentum $\pp=\pp(t)$,
position $\rr=\rr(t)$, and energy $E=E(\pp,\,\rr,\,t)$ given by
Eq.~(\ref{eq:E}). Using the method developed in
Refs.~\cite{Leggett,Yip}, Eq.~(\ref{eq:E}) can be considered as an
effective, semi-classical
Hamiltonian yielding the following equations of motion:
\begin{equation}
\dot{\rr}=\frac{\partial E(\pp,\,\rr,\,t)}{\partial \pp}
=\frac{\ep}{\sqrt{\ep^2+\Delta_0^2}} \frac{\pp}{m^*} + \vv_s(\rr,\,t),
\label{eq:hamilton1}
\end{equation}
\begin{equation}
\dot{\pp}=-\frac{\partial E(\pp,\,\rr\,t)}{\partial \rr}
=-\frac{\partial}{\partial \rr}[\pp \cdot \vv_s(\rr,\,t)].
\label{eq:hamilton2}
\end{equation}
Introducing the non-dimensional variables~\cite{Nugzar_2}
\begin{equation}
\RR=\frac{\rr}{\xi_0}, \, \VV_s=\frac{\xi_0}{\kappa}\vv_s, \,
\tau=\frac{t}{t_0}, \, {\bf\Pi}=\frac{\pp}{\pF}, \, H=\frac{E}{\Delta_0},
\label{nondim}
\end{equation}
where $\RR=(X,\,Y)$, 
$\pF=\sqrt{2m^*\ep}\approx 8.28 \times 10^{-20}~\rm g~ cm/s$
is the Fermi momentum, and
$t_0=\xi_0\pF/\Delta_0\approx2.9\times10^{-6}\,{\rm s}$,
in the non-dimensional form the
Hamiltonian~(\ref{eq:E}) and the equations of motion
(\ref{eq:hamilton1})-(\ref{eq:hamilton2}) become
\begin{equation}
H=\sqrt{\lambda^2({\Pi}^2-1)^2+1}+
m^*m^{-1}\pi^2{\bf\Pi}\cdot\VV_s(\RR,\,\tau),\label{eq:dimhamilt}
\end{equation}
and
\begin{equation}
\dot\RR=
\frac{2\lambda(\Pi^2-1)}{\sqrt{(\Pi^2-1)^2+\lambda^{-2}}}{\bf\Pi}
+\frac{m^*}{m}\pi^2\VV_s(\RR,\,\tau), \label{nondim_R}
\end{equation}
\begin{equation}
\dot{\bf\Pi}=-m^*m^{-1}\pi^2\boldnabla({\bf\Pi}\cdot\VV_s(\RR,\,\tau)),
\label{nondim_Pi}
\end{equation}
where now $\dot{A}\equiv dA/d\tau$, 
and we introduced the non-dimensional parameter
$\lambda=\eF/\Delta_0$. In the following calculations we assume
$\lambda=10^3$.

We solve numerically the equations~(\ref{nondim_R})-(\ref{nondim_Pi})
of ballistic motion of quasiparticles 
in order to calculate the propagation of thermal flux,
generated by the point source of thermal excitations, through the
point vortex gas configurations $1^{\rm o}$, $2^{\rm o}$, and
$3^{\rm o}$ introduced above. The numerical
method is described in our earlier work~\cite{Nugzar_2}.

In turbulence experiments in $^3$He-B, the properties of
the vortex tangle or the gas of vortex rings can be studied by measuring
the heat which is transported by thermal excitations through the
velocity field of the vortices \cite{Bradley-PRL2004} (see also the
review article Ref.~\cite{Fisher2008}). A net flux of excitations (and, hence,
energy) results in the case where there is a (small) temperature
gradient. In this case the heat carried by excitations generated by
the source (and, therefore, incident on the vortex gas) is
\begin{equation}
\delta Q_{inc}=\int_{\Delta_0}^\infty N_F\vF E
\frac{\partial f(E)}{\partial T}\delta T\,dE, \label{Q}
\end{equation}
where $\delta T\ll T$ is a temperature difference between the
source of thermal excitations and the opposite side of the system,
$N_F=m\sqrt{2m^*\eF}/(\pi^2\hbar^3)$
is the density of states at the Fermi energy,
$\vF=\pF/m^*\approx5.48\times10^3\,{\rm cm/s}$ is the Fermi
velocity, and $f(E)$ is the Fermi distribution, which, at the
ultra low temperatures, becomes the Boltzmann
distribution $f(E)=\exp(-E/k_B T)$.

In the case of an isolated vortex, one side of it presents a
potential barrier to oncoming quasiparticles, which are Andreev
reflected as quasiholes almost exactly back to the source. In our
previous work~\cite{Nugzar_1} we found that the Andreev shadow, i.e.
the maximum distance from the vortex core past which a quasiparticle
with the kinetic energy $\ep$ is not Andreev reflected, is 
$S_0=3\pi(\Delta_0/\ep)^2$ in our
dimensionless units. In the subsequent
work~\cite{Nugzar_2} it was shown that, due to partial screening,
the Andreev shadow of simple configurations of several
vortices is not necessarily equal to the sum of shadows of isolated
vortices. Hence we have reasons to expect that
in a gas of point vortices or in a
gas of vortex-antivortex pairs (modelling, respectively, a vortex
tangle or a gas of vortex rings) partial screening
will strongly affect the heat flux carried back to the source by
Andreev reflected quasiparticles.

The numerical simulation of the heat flux reflected by the point
vortex gas was carried out in the rectangular domain. 
At $t=0$ the vortices are randomly distributed
within the square sub domain whose non-dimensional coordinates are
$-5\times10^3\leq X\leq5\times10^3$, 
$-5\times10^3\leq Y\leq5\times10^3$ (the 
size of this domain corresponds to the experimental
estimates~\cite{BradleyPRL2005,Fisher2001,Bradley-PRL2004,Bradley-PRL2006}).
The heat source is located at $(-1\times10^4,\,0)$ so that
the angle, $\varphi$ between the $X$-axis and the beam of
quasiparticles varies
between $-\pi/4$ and $\pi/4$. The flux generated by the source is
modelled by $K=25000$ quasiparticles whose non-dimensional initial 
energies, $H_0$
($1\leq H_0\leq1.688$) and directions, $\varphi_0$ are uniformly 
distributed. Since the source of excitations is located sufficiently
far from the vortices, the initial momentum of quasiparticle whose
initial energy is $H_0$ was calculated from Eq.~(\ref{eq:dimhamilt})
as $\Pi_0=[\lambda^{-1}(H_0^2-1)^{1/2}+1]^{1/2}$. For the Boltzmann
distribution, $f(E)$, introducing $\tilde{E}=E/k_B T$ the integrand
in (\ref{Q}) reduces to the $\Gamma(3)$ distribution which, in our
numerical calculations, was generated by the standard subroutine and
the resulting values were discarded if $\tilde{E}<\Delta_0/k_B T$.
In the typical low temperature experiments,
$\Delta_0/(k_B T)\approx10$; this value was used throughout all
calculations.

A trajectory of each quasiparticle was found by numerical solution of
Eqs.~(\ref{nondim_R})-(\ref{nondim_Pi}) 
for random initial energies (momenta) and directions of
motion. Having identified the trajectories of quasiparticles
that are Andreev reflected (as quasiholes) by the vortex gas or a gas
of vortex-antivortex pairs, for a particular realization, R of the
initial configuration of the vortex gas the reflection coefficient is
then calculated as
\begin{equation}
f_r^{\rm R}=\frac{1}{q_0}\sum_{j=1}^K\alpha_i H_{i0} \quad
{\rm with} \quad q_0=\sum_{j=1}^K H_{i0},
\label{frR}
\end{equation}
where $\alpha_j=1$ if the $j^{\rm th}$ quasiparticle is reflected,
otherwise $\alpha_j=0$, and $K=2.5\times10^4$. This procedure has been
carried out for $N_{\rm R}$ realizations of the initial configuration
of the vortex gas, and the ensemble average reflection coefficient was
calculated as $f_r=\langle f_r^{\rm R}\rangle$ for configurations 
$1^{\rm o}$, $2^{\rm o}$, and 
$3^{\rm o}$. Since the error of ensemble averaging decreases with the
number of realizations as $N_{\rm R}^{-1/2}$, to achieve a reasonable
(few per cent) accuracy we used up to $N_{\rm R}=2.5\times10^4$
realizations.

\begin{figure}[t]
\begin{tabular}[b]{ccc}
\includegraphics[height=0.25\linewidth]{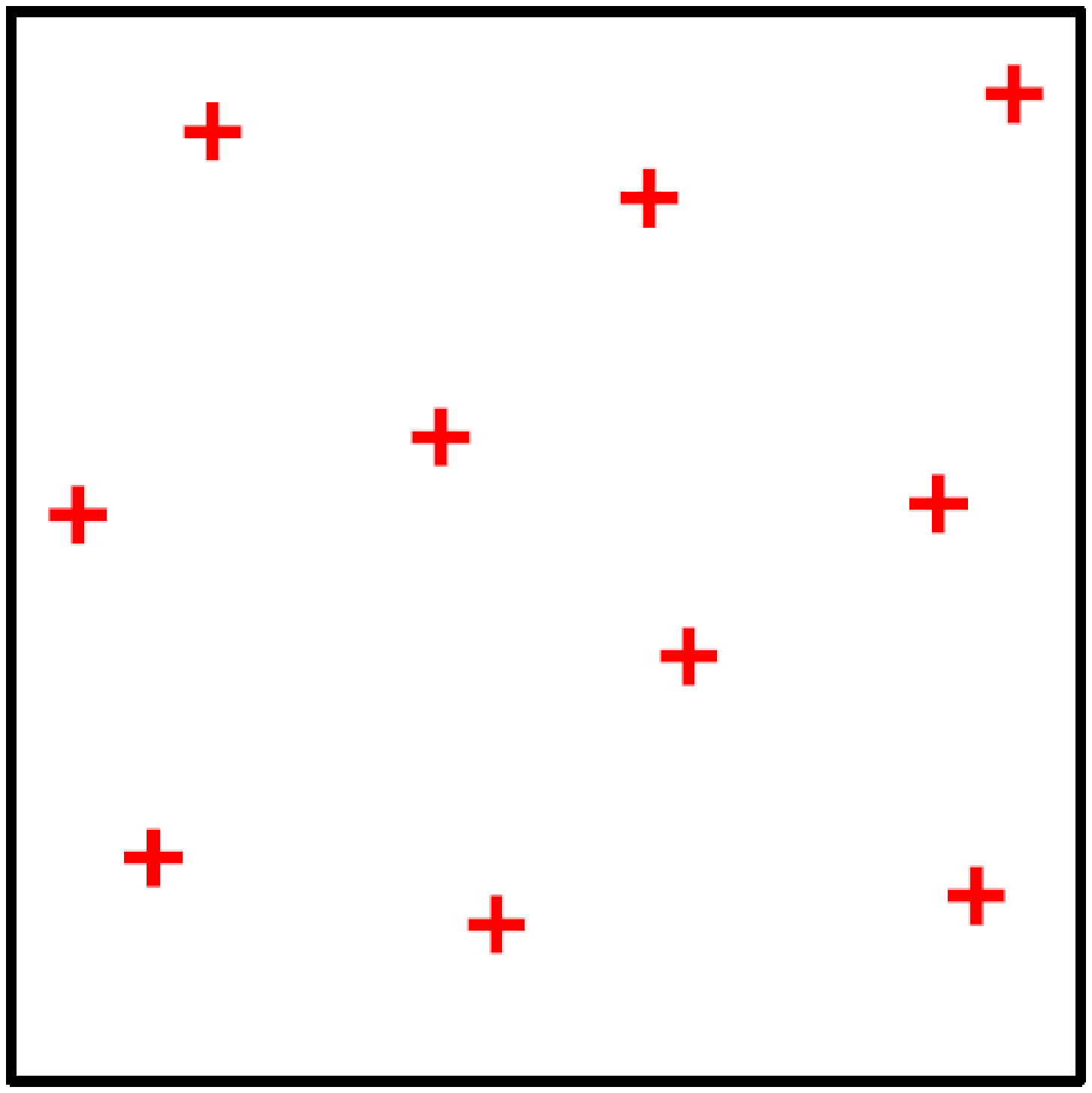}&
\includegraphics[height=0.253\linewidth]{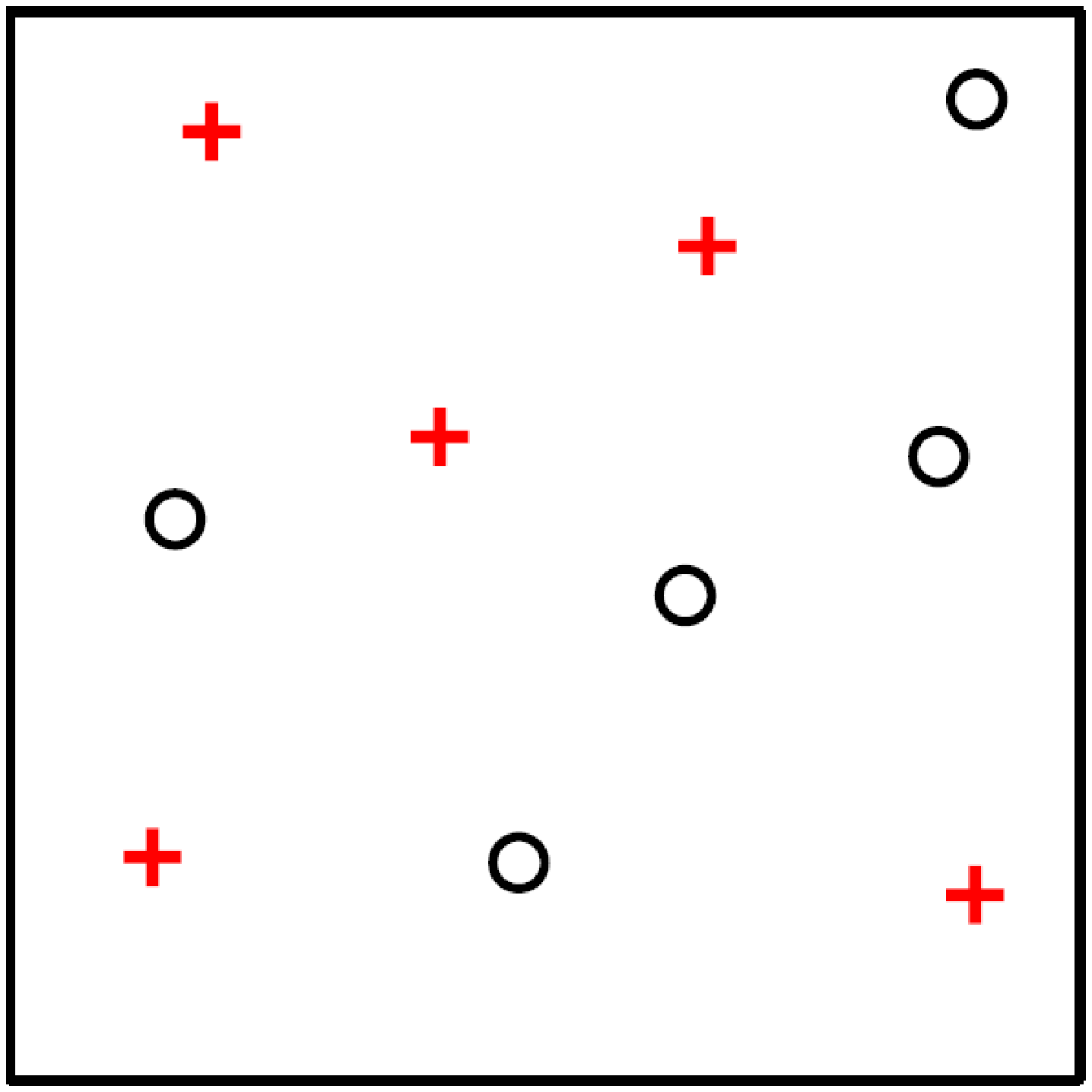}&
\includegraphics[height=0.249\linewidth]{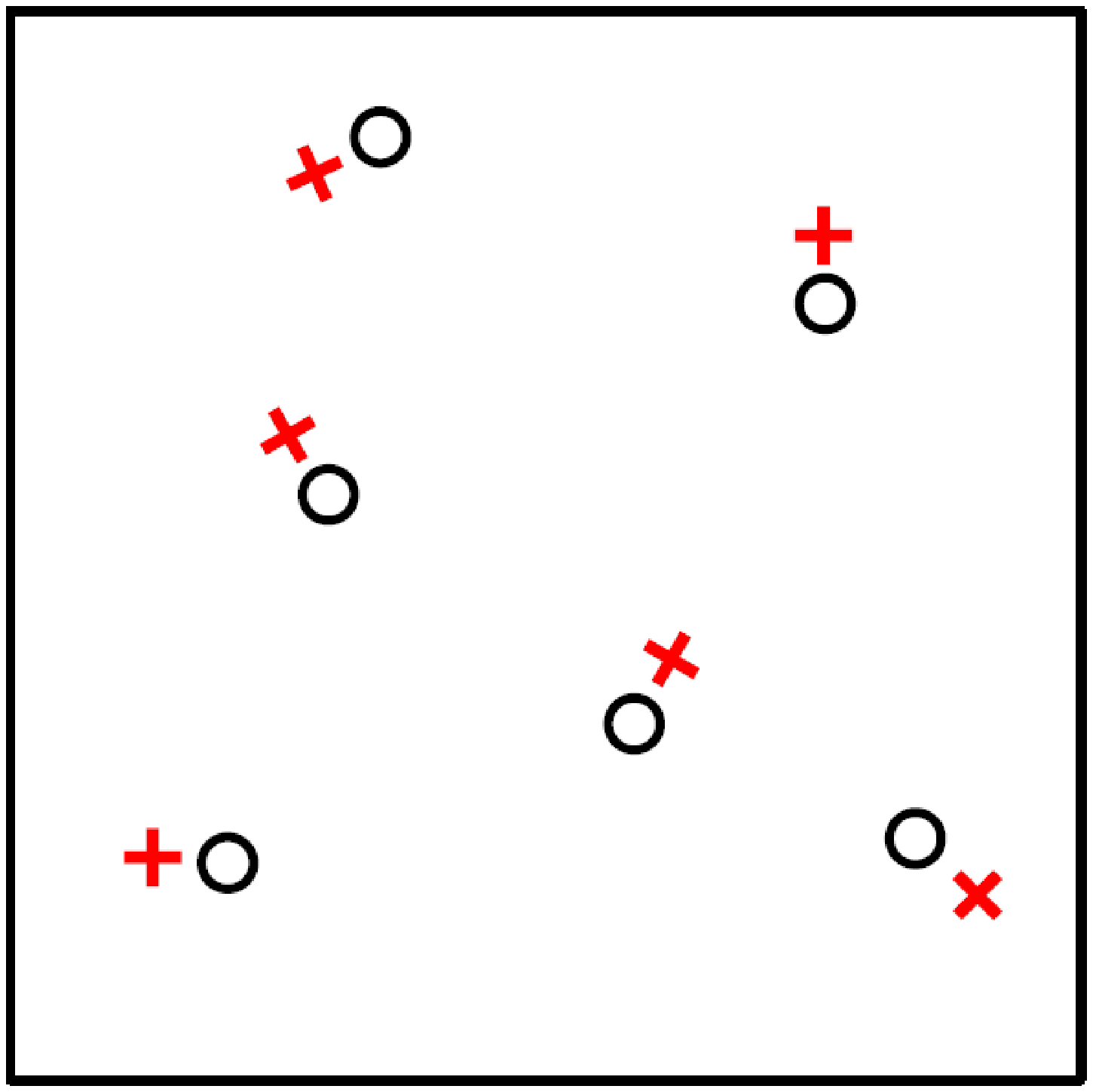}\\
\end{tabular}
\caption{(Color online) Zoomed view of configurations $1^{\rm o}$
(left), $2^{\rm o}$ (center), and $3^{\rm o}$ (right); ${\textbf{+}}$
(red) -- vortices,
${\bf{\bigcirc}}$ (black) --
antivortices.} \label{fig:1}
\end{figure}

\begin{figure}
\centering \epsfig{figure=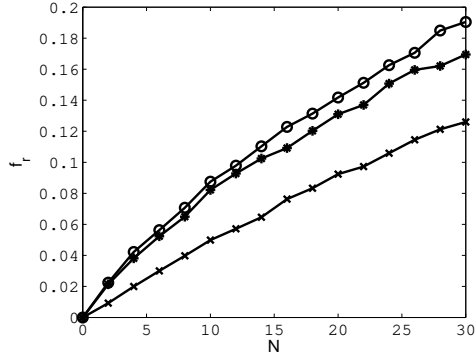,height=2in,angle=0}
\caption{Reflection coefficient as a function of the
total number of vortex points. Lines, from top
to bottom, correspond to
configurations $1^{\rm o}$, $2^{\rm o}$, and the gas of
vortex-antivortex pairs $3^{\rm o}$ with $d=100\xi_0$.}
\label{fig:2}
\end{figure}

At this point it
seems useful to show a zoomed view of each of these configurations,
see Fig.~\ref{fig:1}.
Fig.~\ref{fig:2} shows the reflection coefficient, $f_r$ as
a function of the total number of vortices, $N$ in the system. As
can be seen from this figure, in all cases the reflection
coefficient increases with the total density of the vortex points.
The reflection coefficients in the system $1^{\rm o}$ of vortices
of the same polarity and in the vortex gas $2^{\rm o}$ with zero
net circulation are close, although in the latter case the
partial screening seems to have a slightly bigger effect. However,
in the case where the vortices of opposite polarities form pairs
(`rings'), partial screening plays a far more pronounced r\^{o}le
and the reflection coefficient falls by almost an order of
magnitude (in the calculation illustrated by the bottom
line of Fig.~\ref{fig:2} the dimensional distance between the
vortex points in each pair was assumed $d=100\xi_0$).

\begin{figure}
\centering \epsfig{figure=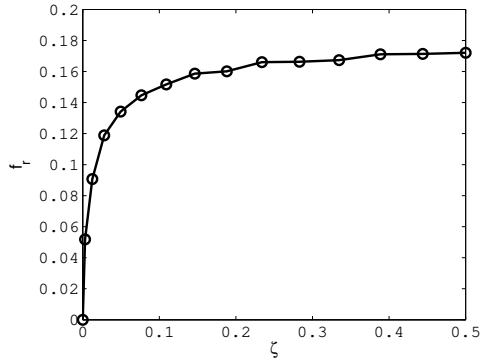,height=2in,angle=0}
\caption{Reflection coefficient, for 13 vortex-antivortex pairs
($N=26$), as a function 
$\zeta=d/a$, where $d$ is the distance between the vortex
and antivortex in a pair, and $a$ is the average distance between the
centers of nearest pairs.}
\label{fig:3}
\end{figure}

As was already mentioned above, of particular interest is a
behaviour of the reflection coefficient during the transition
between configurations $2^{\rm o}$ and $3^{\rm o}$. Such a
behaviour can be characterized by the reflection coefficient,
$f_r$ as a function of the non-dimensional parameter $\zeta$. This
function is illustrated by Fig.~\ref{fig:3} for $N=26$ (for other
values of $N$ the behaviour of $f_r$ with $\zeta$ remains 
qualitatively the same). As can be seen, in the considered example, 
with 13 vortex-antivortex pairs, the increase of $\zeta=d/a$ from
$\zeta=0.01$ (gas of small vortex-antivortex pairs) to $\zeta=0.2$ 
(almost random, disordered point vortex gas) is accompanied by almost
five-fold increase of the reflection coefficient, from 
$f_r\approx3.5\times10^{-2}$ to 0.158, respectively, the latter value
being not very different from that for the random point 
vortex gas ($f_r\approx0.169$). Therefore,
the results shown in Fig.~\ref{fig:3} seem to
indicate that the transition, observed by Bradley 
{\it et al.}~\cite{BradleyPRL2005}, from the gas of vortex rings to the
relatively dense vortex tangle can be detected by the significant 
(nearly an order of magnitude)
increase of the coefficient of reflection of the heat flux carried
by thermal excitations.

The experiment~\cite{BradleyPRL2005} was performed at temperature
$T=0.16~T_c$ and pressure
$P \approx 0~{\rm bar}$ where $T_c$ is the critical temperature.
Quantized vortices were generated by an oscillating grid
and detected by two vibrating wires placed near the grid. 
A beam of quasiparticles illuminated the grid. 
In the presence of vortices a fraction of the
quasiparticles is Andreev reflected, reducing the damping of the
vibrating wire; this damping is caused by the asymmetry of the 
quasiparticles and quasiholes incident upon the wire. 
The transient response of the fractional
change of the damping was measured as a function of the velocity of the
oscillating grid. 
It was found that at high grid velocity ($4.5$ to $7.8~\rm mm/s$)
the fractional reduction of damping was from $0.25$ to $0.4$ and
recovered slowly in  $10$ to $15~\rm s$, 
whereas at small grid velocity ($1.9$ to $2.9~\rm mm/s$) the fractional
reduction of damping ranged from $0.025$ to $0.1$ and recovered
quickly in less than $1~\rm s$. The experimenters suggested
an interpretation based on the recovery time, i.e. that
at high grid velocity quantum turbulence is created which slowly
decay and disperse away. On the contrary,
at small grid velocity the vorticity is in the form
of a gas of vortex rings no larger than $5~\rm \mu m$, which quickly
move away (the translational velocity of a vortex ring is inversely
proportional to its size). Our computed results are consistent with
this interpretation: we found that the reflection coefficient of
a gas of vortex-antivortex points (the two-dimensional equivalent
of a gas of vortex rings) is much less than that of a random vortex
gas (the two-dimensional equivalent of a vortex tangle), which agrees
with the experimental observation.

We are grateful to D.~I.~Bradley, G.~R.~Pickett, S.~N.~Fisher,
W.~F.~Vinen, and also to the participants of the Workshop on Topics in
Quantum Turbulence (Abdus Salam International Centre for Theoretical
Physics, Trieste, March 2009) for stimulating
discussions.

%\vfill
%\eject

\end{document}